%% file: Simulating_carbon_mineralization_at_pore_scale_in_capillary_networks_of_digital_rock.tex
\documentclass[fleqn,10pt]{wlscirep}
\usepackage[utf8]{inputenc}
\usepackage[T1]{fontenc}

\usepackage{tikz} 
\usetikzlibrary{shapes.geometric, arrows, shapes, arrows, calc, arrows.meta} 
\usepackage{graphicx}
\usepackage{color}
\usepackage{longtable}
\usepackage[section]{placeins}
\usepackage{comment}

\usepackage{amsmath,amssymb,calc}
\usepackage{setspace}
\usepackage{ulem}
\usepackage{gensymb}
\usepackage{textcomp}
\usepackage{lineno}
\usepackage{caption, subcaption}
\usepackage{siunitx} 
\usepackage{soul}

\usepackage{eso-pic}

\usepackage{newfloat}
\DeclareFloatingEnvironment[name={Supplementary Figure} ]{suppfigure}
\DeclareFloatingEnvironment[name={Supplementary Table} ]{supptable}

\usepackage{url}

\title{Simulating carbon mineralization at pore scale in capillary networks of digital rock}

\author[1]{David A. Lazo Vasquez}
\author[2,*]{Jaione Tirapu Azpiroz}
\author[2]{Rodrigo Neumann Barros Ferreira}
\author[3]{Ronaldo Giro}
\author[2]{Manuela Fernandes Blanco Rodriguez}
\author[2]{Matheus Esteves Ferreira}
\author[2]{Mathias B. Steiner}
\affil[1]{IBM Research, Rua Tutóia, 1157, São Paulo, SP, 04007-005, Brazil}
\affil[2]{BM Research, Av. República do Chile, 330, RJ, 20031-170, Rio de Janeiro, RJ, Brazil.}
\affil[3]{IBM Research, Rd J Fco Aguirre Proenca Km 9 Sp101, Hortolandia, Sao Paulo, 13186-900, SP, Brazil.}

\affil[*]{jaionet@br.ibm.com}


\begin{abstract}
Predicting the geometrical evolution of the pore space in geological formations due to fluid-solid interactions has applications in reservoir engineering, oil recovery, and geological storage of carbon dioxide. However, modeling frameworks that combine fluid flow with physical and chemical processes at a rock's pore scale are scarce. Here, we report a method for modeling a rock's pore space as a network of connected capillaries and to simulate the capillary diameter modifications caused by reactive flow processes. Specifically, we model mineral erosion, deposition, dissolution, and precipitation processes by solving the transport equations iteratively, computing diameter changes within each capillary of the network simultaneously. Our automated modeling framework enables simulations on digital rock samples as large as (1.125mm)$^3$ with 125$\times 10^6$ voxels within seconds of CPU time per iteration. As an application of the computational method, we have simulated brine injection and calcium carbonate precipitation in sandstone. For quantitatively comparing simulation results obtained with models predicting either a constant or a flow-rate dependent precipitation, we track the time-dependent capillary diameter distribution as well as the permeability of the connected pore space. For validation and reuse, we have made the automated simulation workflow, the reactive flow model library, and the digital rock samples available in public repositories. 
\end{abstract}

\begin{document}

\flushbottom
\maketitle
%
%


\section*{Introduction}\label{sec:intro}

Carbon dioxide (CO$_{2}$) storage in porous sedimentary rocks is a promising technique to reduce the concentration of anthropogenic greenhouse gas emissions in the atmosphere \cite{ali2022recent}. Once  injected into the pore space of the rock, CO$_{2}$ can become permanently trapped following various trapping mechanisms \cite{RABIU2017461}; among them, mineral trapping is a predominant process in basaltic and sedimentary formation \cite{agartan2015experimental, al2017influence}. It occurs when dissolved CO$_{2}$ reacts with brine or saltwater to form carbonic acid and, over time, this carbonic acid reacts with minerals present in the geological formation (e.g., magnesium and calcium silicate) leading to the precipitation of carbonates \cite{wang2015effects}. 
Tracking the impact that these pore-scale processes have in the rock properties requires computing the spatial-temporal evolution of the pore space geometry under the combined effect of the fluid flow and the underlying physical and chemical mechanisms (e.g., erosion, deposition, dissolution, precipitation, and swelling)\cite{matias2021flow, jager2017channelization, molins2012investigation, molins2014pore, noiriel2012upscaling, molins2021simulation, nooraiepour2021probabilistic}. 

Displacements of the fluid-solid interface depend on parameters like flow conditions, fluid phase properties, geometric characteristics of the spatial domains, and the effect of physical and chemical processes taking place within the pores. The product of these pore-scale processes is frequently predicted by transport-reaction numerical models that predict the changes in the porous space geometry under the simultaneous effect of flow speed, flow pressure, volume averaged saturation, and concentration\cite{noiriel2012upscaling, molins2012investigation}. 

Direct numerical methods (e.g., volume-of-fluid \cite{maes2021geochemfoam} or Lattice-Boltzmann \cite{nooraiepour2021probabilistic}) applied on high-resolution mesh-like representations of the pore space geometry are common in solving the combined set of equations describing the mass conservation of the primary species, Navier-Stokes description of the flow at pore-scale, and the model for the fluid-solid interface evolution.\cite{maes2021geochemfoam, matias2021flow,jager2017channelization, molins2015reactive,nooraiepour2021probabilistic} 
The result are long simulation times and high computational cost due to the complex simulation domains and large number of variables involved,  becoming impractical for large mesh sizes and when simulating multiple processes with different time scales. 
For example, according to Matias el al.\cite{matias2021flow}, among the physical processes, the rate of erosion can be 7.5X faster than the rate of swelling. Similarly, according to Molins et al.\cite{molins2021simulation}, among the chemical processes, the geometrical changes due to mineral dissolution can be orders of magnitude slower than the geometrical evolution due to precipitation.
Finally, the time scale associated with transport, measured as the distance traveled by a particle flowing in the fluid, is normally faster than that of the evolution of the fluid-solid interface, measured as the material added to or removed from the pore surface within the time of one iteration.\cite{molins2021simulation,LICHTNER1988143}

In this paper, we report a methodology for adapting existing phenomenological surface models of the geometry evolution in porous media to the capillary network (CN) representation of rock samples,\cite{neumann2021scirep} together with a collection of pore-scale process models and formulations relevant to spatial domains and flow conditions inherent in carbon storage processes. 
As illustrated in Fig. \ref{fig:GMM_workflow}, we track the geometry evolution of the pore space modeled as a network of capillaries due to pore-scale processes (i.e., erosion, deposition, mineral dissolution, and mineral precipitation) by solving the transport-reaction equations iteratively. Depending on the fluid phasic properties, spatial domain characteristics, initial and boundary conditions, the \textit{Flow Simulator} solves for the pressure and flow rate fields at each point in the network. The \textit{Fluid-Solid Process Simulator} then computes the volume gained or lost within each capillary due to each pore-scale process under study, and the resulting change in diameter. Finally, the \textit{Network Geometry Modifier} adjusts the CN representation of the sample accordingly in preparation for the next iteration.
Our combined transport-reaction iterative methodology follows from the procedures in Matias et al.\cite{matias2021flow} and Molins et al.\cite{molins2021simulation}. During each iteration, time-independent transport equations are solved for single-phase flow within the framework of the CN. After extracting flow properties within each capillary of the network, the influence over the spatial domain of pore-scale processes is estimated using phenomenological correlations frequently employed for particle dynamics expressions \cite{bonelli2006modelling, jager2017channelization, matias2021flow} and affinity models of surface-controlled crystallization\cite{noiriel2012upscaling, molins2021simulation}.
A collection of fluid-solid process models and parameters gathered from the literature, to calculate the changes in capillary diameter, is available in section \textit{Model Reaction Library} of the Supplementary Information.

We have applied the methodology to simulate mineral precipitation of calcium carbonate in brine, evaluate its effect on the rock geometrical properties, track the clogging of inlet pores and estimate the volume precipitated and stored within the sample. We have analyzed the influence of using two different correlations and varying reaction parameters on a Sandstone rock sample as an approach to predicting optimum field conditions for carbon storage in reservoirs.

\section*{Methods}\label{sec:methods}

\subsection*{Flow simulations workflow}\label{sec:flowsim}

\begin{figure}[ht]
\centering
\includegraphics[width=\linewidth]{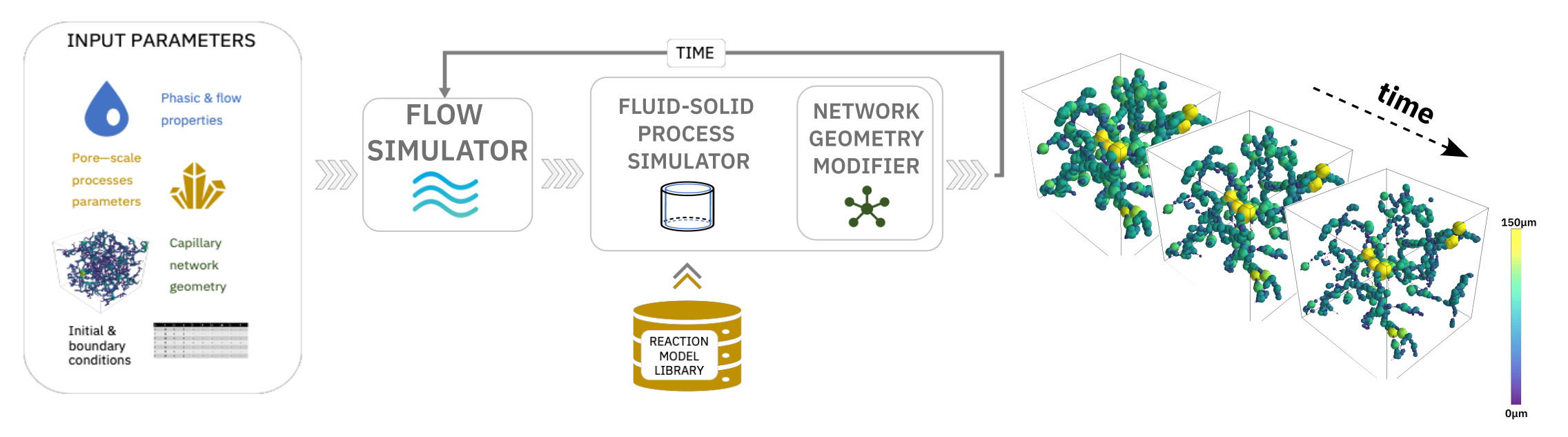}
\caption{Methodology for predicting, within a capillary network representation framework, the spatio-temporal geometry evolution of the porous structure due to fluid-solid interactions by estimating changes in the capillary diameter through an iterative computation.}
\label{fig:GMM_workflow}
\end{figure}

In our work, we use a capillary network, a sparse graph model, as a representation of the pore space of a rock sample\cite{neumann2021scirep}. Starting from a high-resolution three-dimensional digital image of a rock sample obtained through computerized x-ray micro-tomography, we model the rock pore space as a network of connected capillaries or short (one-voxel long) cylinders, and with spatially varying diameter to match the local geometry. Applied to the entire sample, this Capillary Network Model provides an accurate representation of the rock porous geometry as seen in Fig. \ref{fig:GMM_workflow} displaying the CNM representation of a small Berea Sandstone sample with 1,000,000 (100\textsuperscript{3}) voxels, each representing a physical region of (2.25 \textmu m)\textsuperscript{3}, converted to a network of roughly 2,000 capillaries displayed employing a color-coded diameter scale. 

Capillary network representations allow performing both single and two-phase fluid flow simulations with lower computational demands than their mesh- or lattice-based counterparts. Flow simulations assume laminar flow within each capillary to relate flow and pressure, followed by conservation of mass at each network node, to build a large system of coupled equations in sparse matrix form  \cite{neumann2021scirep, TirapuGHGT}. The solution of this system of equations allows determining flow properties like pressure at each node in the capillary network, or flow rate within each capillary, with a high level of geometrical accuracy. Bulk flow properties like permeability are then determined from those fields and can be directly compared to experimentally measured values to benchmark the accuracy of the simulator \cite{neumann2021scirep,Esteves2023}. 
In particular, we ran simulations on samples as large as (1.125mm)$^3$, digitized with 125,000,000 voxels (500 voxels per side of the cube) before converting into a capillary network with 330,000 capillaries, in seconds per iteration on a 4-thread CPU, requiring between 6 hours to 2 days to reach convergence, depending on the parameters of the reaction model. This is significantly faster than equivalently sized simulations using mesh-based simulations like OpenFOAM with 2,300,000 cells requiring the equivalent of 31 CPU-days\cite{maes2021geochemfoam}.

\subsection*{Pore-scale process models}\label{sec:reactions}

The impact of pore-scale processes like erosion, deposition, dissolution, or precipitation within the rock pore space can be described in terms of changes to the diameter of the cylindrical capillaries. This method is frequently applied in areas like soil modeling\cite{kirk2015simple} or channelization in porous media simulations\cite{matias2021flow, jager2017channelization}. As illustrated in Fig. \ref{fig:GMMmethod}a, for each pore-scale process, the associated accumulation or removal of material from the surface modifies the diameter of the cylinders by an amount that may depend on the local fluid phase properties, flow and rock surface conditions, reaction parameters and the pore geometry\cite{molins2017mineralogical, maes2021geochemfoam, matias2021flow}. 
We use phenomenological models from the literature, anchored in physical and chemical studies.
Erosion and deposition are described using particle dynamics expressions\cite{bouddour1996erosion, jager2017channelization, jager2018erosion, fama2020simple, matias2021flow}. The rates and thresholds corresponding to physical processes involving calcium carbonate and brine are defined by the Erosion and Deposition Law in Matias et al.\cite{matias2021flow}, Jager et al.\cite{jager2017channelization}, and Bonelli et al.\cite{bonelli2006modelling}. Mineral dissolution and precipitation rates of calcium carbonate in brine are calculated with phenomenological correlations based on affinity models of surface-controlled crystallization. Dissolution rates are defined by the correlation of Molins et al.\cite{molins2021simulation}. Precipitation rates are defined by the correlations of Lasaga et al.\cite{lasaga1981kinetics} and Noiriel et al.\cite{noiriel2012upscaling}. 
In this work we describe mineral precipitation in porous media as a deterministic process, applying reaction rates uniformly to all sites in the network without considering possible nucleation patterns arising from the probabilistic nature of the process\cite{nooraiepour2021probabilistic}.

Several diameter-scaling models and formulations for various pore-scale phenomena as collected from the literature are included in the \textit{Supplementary Information} and \textit{Code Availability} sections.

\subsection*{Fluid-solid interface dynamics}\label{sec:fluid-solid_interface}

\begin{figure}[ht]
\centering
\includegraphics[width=\linewidth]{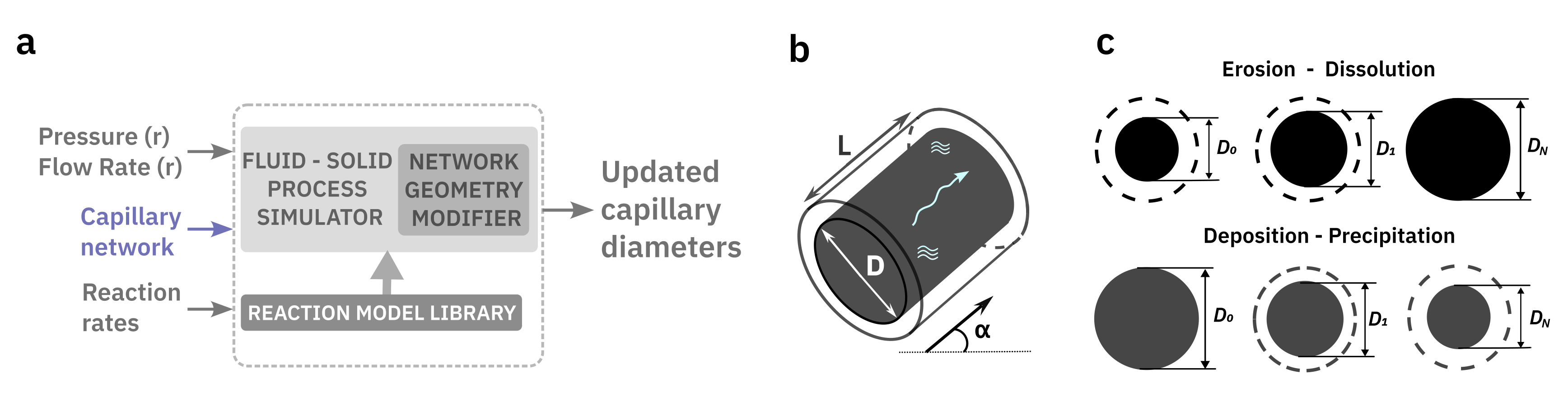}
\caption{a) Summary of inputs and outputs of the reaction model used in the \textit{fluid-solid process simulator \st{module}}. b) Capillary representation in space with diameter and length $D$ and length $L$, respectively, and inclination $\alpha$. c) Capillary diameter evolution due to pore-scale processes during $N$ simulation time steps causes an increase or decrease in the diameter depending on the process in effect. }
\label{fig:GMMmethod}
\end{figure}

Following the workflow in Fig. \ref{fig:GMM_workflow}, the flow simulator takes as input the capillary network model of the rock sample, fluid properties (i.e. viscosity, density) and simulation conditions (i.e. absolute pressure, temperature, and driving pressure gradient across the sample) and returns the pressure and flow rate fields at each node in the network under the assumptions of single-phase, time-independent flow. Turning to Figure \ref{fig:GMMmethod}a, these fields, together with the current CN model geometry, flow conditions and process properties, are transferred to the \textit{Fluid-solid Process Simulator}. 
This step tracks the evolution of the fluid-solid interface geometry within the capillary network at each simulation iteration, computing the joined effect of the fluid flow and the pore-scale processes on the surface of the rock. 

Each capillary within the network is defined as a hollow cylinder of length $L$, diameter $D$, and inclination $\alpha$ with respect to the horizontal, as illustrated in Fig. \ref{fig:GMMmethod}b. 
The simulations neglect spatial heterogeneity at scales smaller than the size of the original digital rock discretization grid\cite{li2006upscaling}. The reaction area per capillary is thus assumed to be uniform, i.e., crystallization and material accumulation (or removal) is uniformly distributed along the length and perimeter of each capillary, and the change in capillary diameter is computed as the average effect of the processes acting at each temporal iteration. 
The interconnected nature of a capillary network in a confined space supports the use of bulk rather than surface-based properties and rates as inputs.
The time step is selected based on the minimum interval in which any reaction produces a noticeable change in diameter relative to the original discretization grid.\cite{li2006upscaling}  
Considering the size of the rock samples under study ($\sim 1.25mm$ side length), and time-independent flow, it is reasonable to assume uniform distribution of mineral concentration (i.e., no concentration gradients or gradients in reaction rates within the network) and constant mineral reaction rates within each capillary during each iteration that only change from one capillary to the next.\cite{jager2017channelization} 
The liquid phase is assumed incompressible, the pressure drop constant and the porous medium consolidated and incompressible \cite{jager2017channelization}. Other processes, such as fluidization \cite{johnsen2008decompaction} and channelization \cite{jager2017channelization, mahadevan2012flow, kudrolli2016evolution} are not taken into account.

The \textit{Network Geometry Modifier} adjusts the diameter of each capillary in the network according to the displacement computed in the previous step, generating a new CNM to be used as geometrical input for the next iteration.
The system iterates this simulation workflow until either reaching a pre-determined number of iterations, or all capillary diameters reach minimum or maximum pre-determined values. For mineral precipitation and solid particle deposition simulations, the minimum capillary diameter is half a voxel. For mineral dissolution and solid particle erosion, simulations stop when the void space reaches the rock sample size.

\subsection*{Pore-scale processes thresholds}\label{sec:onset}

Some of the processes considered in the simulations may display an onset threshold, a short delay before showing noticeable spatial impact.
These minimum reaction times are based on empirically observed thresholds and influence the temporal lapse as well as the processes considered when computing the capillary diameter adjustment per iteration. 
The thresholds for the onset of the chemical processes are based on affinity models of surface-controlled crystallization following the procedure of Molins, S. et al. \cite{molins2021simulation} and Noiriel, C., et al.\cite{noiriel2012upscaling}, in which, after a short delay, the average reaction rate increases sharply with time, leading to significant morphology changes. For example, dissolution processes may require an interval of 2 seconds \cite{molins2021simulation} to initiate, while precipitation may require up to 1,200 seconds \cite{noiriel2012upscaling}. 
In physical processes, the thresholds are defined by the calculation of an erosion or deposition timescale, based on particle dynamics expressions (i.e., the Erosion and Deposition Law in Jager R. et al.\cite{jager2017channelization}).
The iteration time step is chosen as the minimum interval necessary to produce noticeable changes among the various processes being simulated.

\subsection*{Multiple pore-scale processes}\label{sec:multiple}

Each capillary within the network can suffer various geometry-altering processes simultaneously\cite{matias2021flow, noiriel2012upscaling} and the effect of each contributing process may depend, in addition to the process governing rates, on factors like the local geometry and flow conditions (e.g. pressure or flow rate can influence the onset and rate of erosion). 
Under the assumption of a sharp and impermeable reactive area equal to the physical fluid-solid interfacial surface\cite{molins2012investigation}, we can also assume a constant reaction rates during each time step\cite{molins2017mineralogical}. Further assuming that geometrical changes to the pore space have a negligible effect on the flow and transport solutions during each iteration, we can compute the changes to the geometry due to each process separately.
The diameter of each capillary in the network is thus computed as the average variation due to each contributing process as follows:
\begin{equation}
    \hspace*{5cm}\mathbf{D}  \left ( t_N \right ) = \mathbf{D} \left ( t_{N-1} \right ) + \sum_{\theta}^{}\Delta \mathbf{D}_\theta \left ( t_N \right )
\end{equation}
where $\mathbf{D}\left ( t_N \right )$ represents the vector containing the $M$ number of capillary diameters at time step N, $\Delta \mathbf{D}_{\theta}$ is the diameter variation vector due to the effect of pore-scale process $\theta \in \left\{ \textup{erosion, deposition, dissolution, precipitation} \right\}$. 
Solid particle erosion and mineral dissolution produce positive feedback on the local geometry variation, increasing the diameter of the capillary ($\mathbf{D} \left ( t_{N} \right ) > \mathbf{D} \left ( t_{N-1} \right )$ or $\Delta \mathbf{D}_{er, dis} > 0 $), while solid particle deposition and mineral precipitation generate negative feedback, reducing the diameter with each iteration ($\mathbf{D} \left ( t_{N} \right ) < \mathbf{D} \left ( t_{N-1} \right )$ or $\Delta \mathbf{D}_{dep, ppt} < 0 $). A graphical depiction of the diameter adjustment per process type is displayed in Fig. \ref{fig:GMMmethod}c. 
As a result, the final diameter variation of each capillary within the network may be positive or negative depending on the balance between all contributing pore-scale processes, leading to a local dependency on the geometry and flow evolution.

One challenge when simulating simultaneous reactions arises from the disparate spatial and temporal scales associated with them, often orders of magnitude apart.\cite{noiriel2012upscaling}. 
To account for any temporal disparity, the iteration time step is selected in which at least one of the processes being simulated produces a change in diameter larger than the discretization grid of the original digital rock image.\cite{li2006upscaling}
To improve computation efficiency, after each iteration, only those processes producing change in the diameter larger than the grid resolution are considered when adjusting the diameter. 
Moreover, the simulator modifies the geometry of only those capillaries with adjustments larger than a voxel size, otherwise the diameter remains constant during the iteration. Some processes may require multiple iteration before amounting enough change to be included in the computation.

\section*{Results}\label{sec:results}

We have applied the methodology of Fig. \ref{fig:GMM_workflow} to simulate the effect of precipitation to the porous geometry of a Berea Sandstone rock sample\cite{NeumannDataset2020}.  Using as an example the precipitation kinetics of calcium carbonate, the main component in limestone, a promising rock candidate for CO$_{2}$ sequestration\cite{ARIF2017113}, we assumed the rock geometry to represent a calcium carbonate matrix injected with brine. We further assumed single-phase fluid and constant calcium concentration within the the time interval of one iteration, even if the outlet solution concentrations tend to present a slight decrease with time \cite{noiriel2012upscaling}. We compared the effect of applying two different models of precipitation with varying precipitation rates on various metrics relevant to of mineral storage. 

\subsection*{Effect of precipitation on capillary network properties} 

\begin{figure}[ht]
\centering
\includegraphics[width=\linewidth]{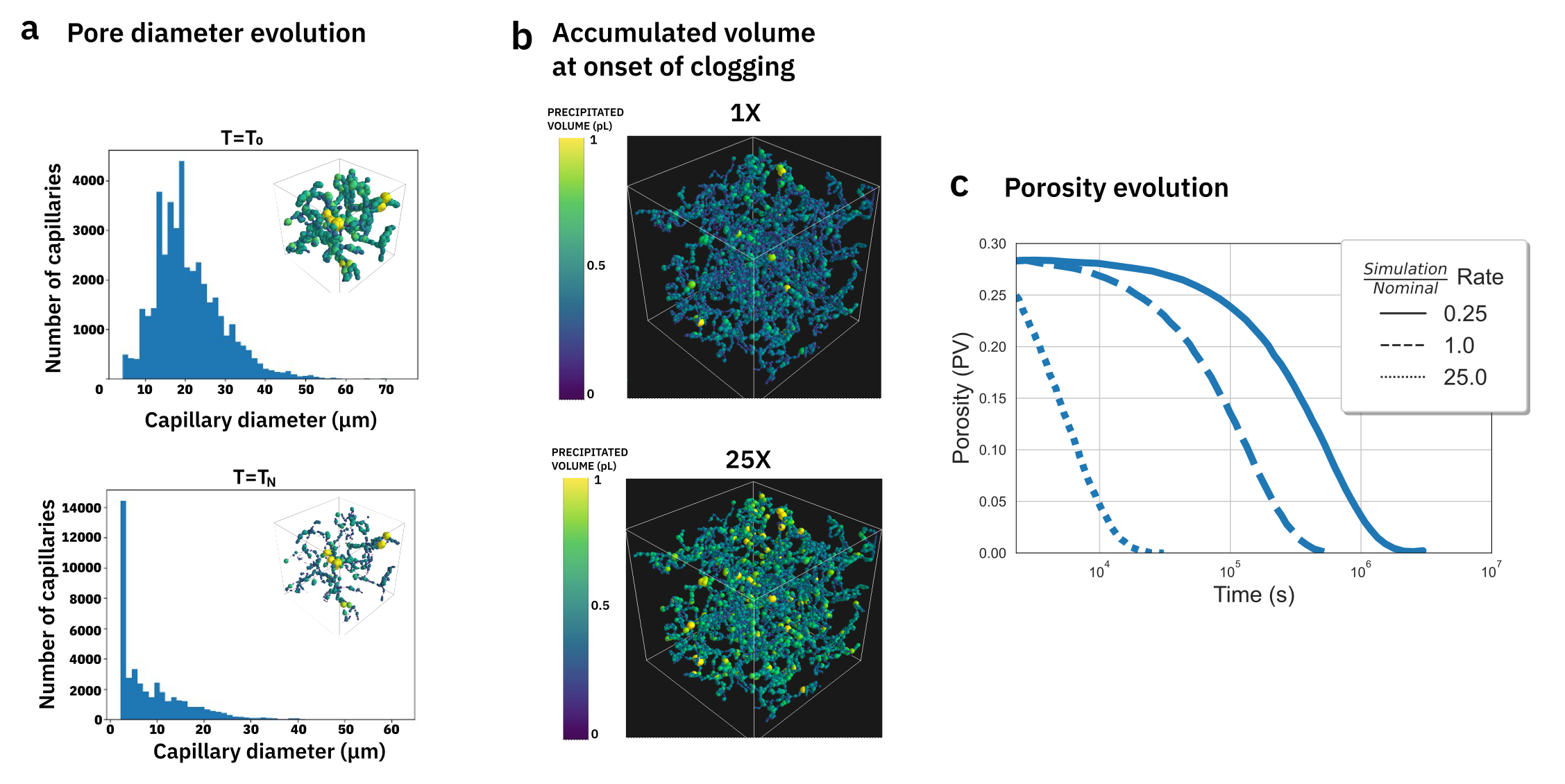}
\caption{a) Evolution of the distribution of capillary diameters on the CNM of the (250 voxels)$^3$ portion of the rock sample as a result of precipitation.  b) Comparison of accumulated precipitated volume at the onset of clogging for two values of precipitation rates, one at the nominal precipitation rate of calcium carbonate in brine and one at a 25-times higher rate. c) Evolution of rock sample porosity as precipitation reduces the pore diameter across the sample at a rate equal to 25 times (25$X$) and 0.25 times (0.25$X$) that of the nominal rate 1$X$ of calcium carbonate in brine. }
\label{fig:GMMresults}
\end{figure}

Fig. \ref{fig:GMMresults}a illustrates an example of the evolution of the distribution of pore diameters as a result of precipitation on a portion of the rock sample with (250 voxels)$^3$ , dimensions of (562.5$\mu$ m)$^3$, about 28\% porosity, modeled by a capillary network of 41720 capillaries. The simulations used a \textit{uniform} precipitation model as discussed in sub-section \textit{Clogging prediction} of the Supplementary Information in relation to Eq. (\ref{eq:correlation1}), in which material accumulation is assumed equally distributed across capillaries. The precipitation rate and other simulation parameters collected in Table \ref{tab:sample} are based on correlations to experimental data of calcium carbonate precipitation in brine\cite{lasaga1981kinetics,noiriel2012upscaling,tang2008sr2+}.
The plot in Fig. \ref{fig:GMMresults}a represents the distribution of  capillary diameters at the beginning of the simulations ($T_0$) and after the $N^{th}$ iteration ($T_N$). The insets in the plot display the resulting CNM of the pore space geometry as the distribution of capillary diameter shifts towards smaller values and the pore space shrinks after a few iterations.

As the capillary diameters gradually decrease as a result of precipitated carbonate, the pores become clogged and can no longer sustain flow, thus rapidly reducing the porosity and permeability of the rock sample. The proportion of clogged pores in the rock sample after several iterations is directly related to the precipitation rate, resulting in larger volumes for higher rates. Fig. \ref{fig:GMMresults}b displays a comparison of accumulated precipitated carbonate volume at the onset of clogging for two values of precipitation rates, one using the nominal precipitation rate of calcium carbonate in brine as in Table \ref{tab:sample}, and one using a rate 25 times higher, clearly showing a larger accumulated volume and thus higher probability of clogging for the 25$X$ rate example. 

Another consequence of precipitated volume accumulation is the reduction in porosity of the sample, as pores shrink and even become clogged. Fig. \ref{fig:GMMresults}c plots the evolution of rock sample porosity as precipitation reduces the pore diameter across the sample at a rate equal to 25 times (25$X$) and 0.25 times (0.25$X$) that of the nominal rate 1$X$ of calcium carbonate in brine. The simulations stop in two scenarios: when the capillaries in contact with the inlet face are fully clogged, or when the geometry changes stop being sufficiently significant.

\subsection*{Comparing precipitation model outcomes} 

In the following, we performed simulations on a $(1125\mu m)^3$ portion of the rock sample of size (500 voxels)$^3$ at a resolution of (2.25 \textmu m)\textsuperscript{3} per voxel. 
The geometrical framework consisted of a capillary network of roughly 330$\times 10^3$ capillaries and 227$\times 10^3$ nodes, and an initial porosity $\phi_{0}$ of 32$\%$.
We studied the effect of applying two different precipitation models. 
The first \textit{uniform} model of the previous section, which considered a constant rate of material accumulation uniformly within all capillaries, using as input the set of experimentally validated surface mineralization parameters from Noiriel, C. et al.\cite{noiriel2012upscaling} and Tang et al.\cite{tang2008sr2+}. 
The second \textit{variable} model, which estimated material accumulation considering a precipitation rate that accounts for the influence of the capillary flow rate. The correlation used for computing precipitation rates (see Eq. 3 in Noiriel et al.\cite{noiriel2012upscaling}) is a function of three local variables: volumetric flow rate, total surface area, and the difference between the concentration of calcium determined in the effluent $Ca_{\textup{out}}$, and the well-mixed stock solution $Ca_{\textup{in}}$. 
The reaction time shown in Table \ref{tab:sample} represents the minimum time interval necessary for noticeable impact to the spatial domain. It was computed prior to the simulations and used as the time interval per iteration.
Other simulation parameters related to rock geometry, flow simulation (i.e., the pressure gradient across the sample inlet and outlet interfaces driving the flow or the time step considered per iteration), and reaction parameters are detailed in Table \ref{tab:sample}. 

We studied the impact of varying precipitation rates on four sets of parameters relevant to the geological sequestration of CO$_{2}$\cite{TirapuGHGT,noiriel2012upscaling}. 
The precipitated volume evolution and precipitation volume per iteration allow to make estimates of the geometry evolution, and storage capacity. The number of clogged capillaries and the time to reach a minimum porosity value give an indication of the ability to effectively permeate the rock pore space and permanently store CO$_2$.
The two models for simulating mineral precipitation and consequent clogging are detailed in the \textit{Reaction Model Library section} of the Supplementary Information.

\begin{longtable}[c]{@{}lllll@{}}
    \caption{Flow, geometry, and pore-scale process parameters for simulating mineral precipitation of calcium carbonate in a 500\textsuperscript{3} voxel size rock sample for the reference simulations.}
    \label{tab:sample}\\
    \toprule
                            & Parameter                     & Symbol    & Value    & Units          \\* \midrule
    \endfirsthead
    \endhead
    \bottomrule
    \endfoot
    \endlastfoot
    Fluid flow \cite{TirapuGHGT, TirapuSPIE23}              & Driving force                 & $F$                       & 10132.5   & N                     \\
                            & Pressure gradient             & $\Delta P$                & 101325    & Pa                    \\* \midrule

    Rock sample \cite{TirapuGHGT, TirapuSPIE23}              &  Side length                 & $L_{r}$                       & 1.125   & mm                     \\
    
    & Number of capillaries         & N                         & 330058      & -                     \\
    
                            & Voxel size                    & $voxel$                   & 2.25$\times 10^{-6}$  & m                     \\* \midrule

    Capillary network \cite{TirapuGHGT, TirapuSPIE23}                 
                            & Initial porosity                      & $\phi$                    & 0.32      & -                     \\

                            & Minimum capillary diameter      & $R_{\textup{min}}$        & 0.5       & voxel                 \\* \midrule

    Mineral precipitation \cite{noiriel2012upscaling, tang2008sr2+}   & Precipitation rate constant       & $\kappa_{\textup{ppt}}$   & 4.68$\times 10^{-7}$  & mol m$^{-2}$ s$^{-1}$ \\
                            & Reaction time                 & $t_{ppt}$                 & 1200      & s                     \\
                            & Saturation index              & $\Omega$                  & -0.198     &    -                   \\
                            & Calcium concentration variation        & $\Delta$ Ca                  & 1.336      &    mmol$\cdot$l$^{-1}$                   \\
                            & m Coefficient              & m                  & 1.00      &    -                   \\
                            & n Coefficient              & n                  & 1.00      &    -                   \\
                            & Precipitation rate            & $r_{ppt}$                  & 5.607$\times 10^{-7}$   &   mol m$^{-2}$ m $^{-1}$                    \\* \midrule

    Phasic properties       & Liquid density                & $\rho_l$                  & 1100      & kg m $^{-3}$          \\
                            & Liquid dynamic viscosity      & $\mu_l$                   & 1.002e-3  & Pa s                  \\
                            & Solid density                 & $\rho_s$                  & 2710      & kg m $^{-3}$          \\
                            & Solid molar mass              & $M_s$                     & 0.1       & kg mol$^{-1}$          \\
                            & Liquid temperature              & $T$                     & 300       & K
                            \\* \bottomrule

\end{longtable}

\begin{figure}[ht]
\centering
\includegraphics[width=0.8\linewidth]{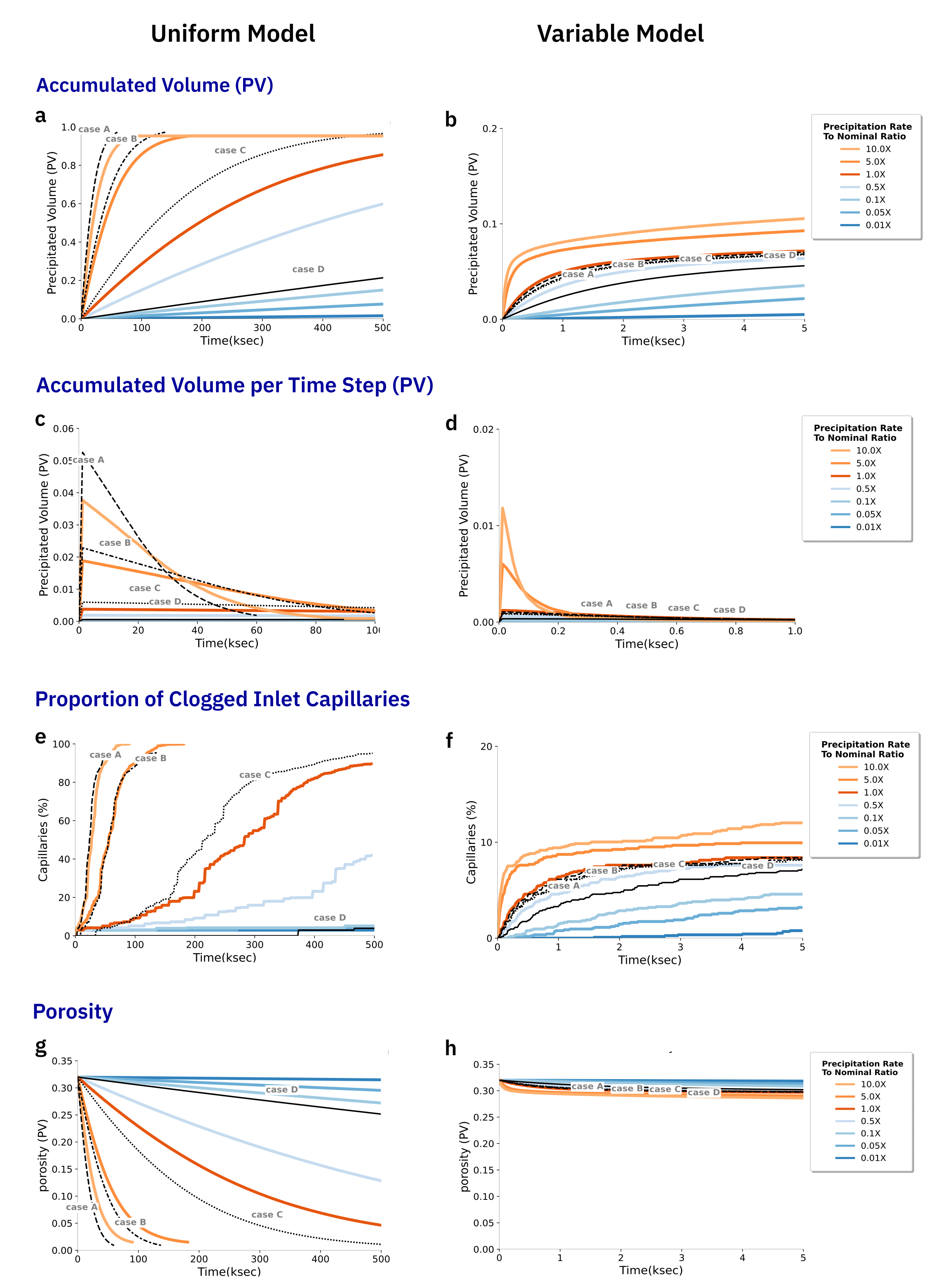}
\caption{Simulation results comparing uniform model in Eq. (\ref{eq:correlation1}) vs the flow-dependent variable model Eq. (\ref{eq:correlation2}). a) Total precipitation volume stored in pore volume (PV) as a function of time computed using uniform precipitation model and b) using the variable model. c) Accumulated precipitation volume at each time step using uniform precipitation model and d) using the variable model. e) Proportion of clogged inlet capillaries as a function of time computed using  uniform precipitation model and f) using the variable model. g) Sample porosity evolution as a result of precipitation as computed with the uniform model and h) with the variable model.}
\label{fig:model_comparison}
\end{figure}

We compared the results of simulating precipitation inside the pore space of the rock sample with both models. We computed the volume of accumulated material stored in each capillary in the network after each iteration $N$ from the difference in the volume of a cylinder of diameter D$_{N}$ minus the volume of a cylinder with diameter D$_{N-1}$. Adding the increments from all capillaries provides the incremental precipitated volume per iteration, and adding all increments for all iterations gives the total accumulated volume stored. Fig \ref{fig:model_comparison}a displays the total precipitation volume as a function of time stored in pore volume (PV) computed using uniform precipitation model of Eq. (\ref{eq:correlation1}), while Fig \ref{fig:model_comparison}b displays the results using the variable flow-rate dependent model of Eq. (\ref{eq:correlation2}). It is observed that the time scale of the two models differ by about a couple of orders of magnitude when applying the same precipitation rate to the formulations, resulting in faster initial precipitation and process saturation as seen in Fig. \ref{fig:model_comparison}b and an overall smaller volume of precipitation with the variable model. 
Fig \ref{fig:model_comparison}c and \ref{fig:model_comparison}d plot the incremental precipitated volume per iteration with the uniform and variable models, respectively.

We analyzed the effect of the precipitation rate on mineral trapping by studying three scenarios. A reference case, denoted as 1.0X, with rate $r_{ppt}$ based on precipitation of calcium carbonate in brine (see Table \ref{tab:sample}),  a regime of five and ten times higher rates denoted as 5.0X and 10.0X, and a regime of lower rates with values 50$\%$, 10$\%$, 5$\%$ and 1$\%$ of the nominal precipitation rate, denoted as 0.5X, 0.1X, 0.05X and 0.01X in Fig \ref{fig:model_comparison}, respectively.  As the precipitation rates increase, the time for reaching a accumulation plateau in Fig \ref{fig:model_comparison}a and \ref{fig:model_comparison}b, and consequently converge to a porosity value, decreases as well. 
Fig \ref{fig:model_comparison}e and \ref{fig:model_comparison}f display the percentage of clogged capillaries at the inlet due to mineral precipitation as predicted by both models for the varying rates of precipitation. As expected, the higher the mineral precipitation rates, the faster clogging of the capillaries occurs. As the capillary diameters decrease with time, in order to maintain numerical integrity of the capillary network, once a capillary reaches a minimum diameter of half a voxel size it is considered clogged. This allows running the \textit{Flow Simulator} in the new geometry and still calculate flow rate, speed, and pressure distribution within the entire network needed for executing the \textit{Fluid-solid process simulator}. 
Finally, Fig \ref{fig:model_comparison}g and \ref{fig:model_comparison}h display the evolution of the sample's porosity as the precipitation volume increases and the pore diameters decrease for the varying rates of precipitation according to the uniform model and the variable model, respectively. \\

\section*{Discussion}\label{sec:discussion}


Together with the range of precipitation rates obtained by scaling the nominal 1.0$\times r_{ppt}$ (1X) from Table \ref{tab:sample}, we overlaid the results of applying a few representative surface precipitation rates from the literature  (see Table \ref{tab:literature_rates}). The precipitation rates in Table \ref{tab:literature_rates}  \cite{tang2008sr2+} were calculated from Eq. (17) in Noiriel et al.\cite{noiriel2012upscaling}, based on parameters from Tang et al.\cite{tang2008sr2+}, obtained through affinity models of surface-controlled crystallization, assuming that the bulk solution is immediately adjacent to the calcium carbonate surface. Cases labeled as A, B, C and D in Fig \ref{fig:model_comparison} cover the range of rates observed in their work.

Substantial differences are observed between the uniform and variable models of precipitation (i.e. left and right plots in Fig. \ref{fig:model_comparison}, respectively).
According to the results displayed in Fig. \ref{fig:model_comparison}a for accumulated precipitation using the uniform model, it is possible to fill the entire pore space of the rock with precipitated mineral, reaching complete saturation faster for higher rates. The accumulated precipitation results using the variable model, on the other hand, predict a maximum precipitation volume of the order of 10\% of the pore space regardless of rate value as seen in Fig. \ref{fig:model_comparison}b. Fig. \ref{fig:model_comparison}c and \ref{fig:model_comparison}d display the precipitation added in each iteration according to both models, showing a fast accumulation within the first simulation iterations, followed by gradually smaller increments. The flow-rate dependent variable model in Fig.\ref{fig:model_comparison}d seems to produce a faster decrease in added precipitated volume per time step after the first initial iterations as compared to the results obtained using the uniform model of Fig. \ref{fig:model_comparison}c, where the decrease seems influenced only by the available space.

Turning to the results for the accumulated precipitation in Fig \ref{fig:model_comparison}b and \ref{fig:model_comparison}d, or the proportion of clogged capillaries in Fig. \ref{fig:model_comparison}f, 
computed applying the parameters of cases A, B, C and D in Table \ref{tab:literature_rates} to the variable model, they do not appear to follow the linear trend of larger precipitation accumulation for higher rates observed with the uniform model in Fig \ref{fig:model_comparison}a, \ref{fig:model_comparison}c and \ref{fig:model_comparison}e.
This behavior is explained by considering that the simulations using a rate directly scaled from the nominal 1.0X, also assume constant calcium variation within the sample, while examples A, B, C and D, employ concentrations measured in the reported experiments \cite{tang2008sr2+,noiriel2012upscaling}, the concentration applied to the rest of the simulations is directly scaled from the 1.0X nominal example, leading to the different trends in Fig \ref{fig:model_comparison}b, \ref{fig:model_comparison}d and \ref{fig:model_comparison}f.

\begin{longtable}[c]{lcccc}
\caption{Examples of calcite precipitation parameters at 25$^\circ$ adapted for surface  from Tang et al.\cite{tang2008sr2+} recalculated with Eq. (17) in Noiriel et al. \cite{noiriel2012upscaling} as a function of saturation state}
\label{tab:literature_rates}\\
\hline
Id. & $\log \Omega $  & $ \dot{r}_{ppt}$ $\left ( \textup{mol} \; \textup{m}^{-2} \textup{s}^{-1} \right )$ & Ca$_{\textup{out}}$ - Ca$_{\textup{in}}$ $\left ( \textup{mmol} \; \textup{l}^{-1} \right )$ & Label \\ \hline
\endfirsthead
\endhead
\hline
\endfoot
\endlastfoot
%
Tang-13 & 1.108 & \num{3.771e-6} & 0.73 & B \\
Tang-14 & 0.931 & \num{8.868e-7} & 1.16 & C \\
Tang-15 & 0.660 & \num{8.171e-8} & 0.99 & D \\
Tang-21 & 1.233 & \num{1.010e-5} & 0.38 & A \\
\hline
\end{longtable}

To understand the validity of the simulation method and the precipitation models, we compare the results in Fig \ref{fig:model_comparison} with those from the stirred reactor experiments in Noiriel et al. \cite{noiriel2012upscaling}. While the work in Noiriel et al. is not directly comparable, it provides experimental evidence for precipitated carbonate volume within the confined spaces of a synthetic porous structure at the microscopic scale, and a rough estimate for the expected precipitation volume range computed by the models considered in this study. 
The precipitation parameters of the study include an initial specific surface area of 0.012m$^{2}$ $g^{-1}$ and a 1 g initial mass of calcite spar as seed crystals.
The experiments were carried out in 70 mL reactors at a total flow rate of 0.5 ml h$^{-1}$ flowing within spaces of the order of 100$\mu m$, with separate streams of NaHCO$_3$ and CaCl$_2$ stock solutions mixed directly in the reactor through two separate injection ports. 
In our work, we assumed a calcium carbonate matrix injected with brine and we simulated the flow within the pore geometry extracted from a Berea Sandstone rock sample of size $(562.5\mu m)^3$ with pore diameters distributed between 20$\mu m$ and 60$\mu m$,  a maximum capillary flow-rate of 0.0882 nL s$^{-1}$ (0.00032 ml h$^{-1}$), and an initial surface reaction area of $6.93\times 10^{-5} m^2$.
According to the experimentally measured precipitation distribution along the length of the reactor in Noiriel et al. \cite{noiriel2012upscaling}, a mean accumulated volume around 2\% or 3\% could be expected, closer to the estimates of the \textit{variable} precipitation model, suggesting the need to considers local phenomena (i.e., concentration distribution, capillary flow rate) to improve the accuracy of the computations.

The discrepancies between the two model types can be attributed to two main reasons: 
(i) the \textit{uniform} model does not consider the effect of the capillary flow rate on the precipitation rate, which is assumed to be a constant, leading to a uniform material accumulation within the network. 
(ii) The \textit{variable} model reveals the impact of considering changes in the distribution of flow rate and other parameters within the network as capillaries start to clog after the initial conditions. 
(iii) The \textit{variable} model also allows adjusting the calcium concentration gradient $\Delta Ca$ applied on the capillaries with data from experiments or the literature, with noticeable influence as observed in the results of Fig. \ref{fig:model_comparison}f, even when assuming a constant concentration across the sample.

As seen in Fig \ref{fig:model_comparison}e and \ref{fig:model_comparison}f, the speed at which the inlet capillaries clog increases as the mineral precipitation rates increase, presumably blocking further injection of carbon dioxide rich brine and, consequently, hindering further mineral storage. 
Given that the spatial domains represent large and complex geometries for reservoir engineering, oil recovery, and carbon dioxide geological sequestration, the ideal mineral precipitation and storage process should develop uniformly and smoothly, occupying all the available pore space.  
The prediction of the onset of each pore-scale process, i.e., specifically the capillaries where a reaction such as precipitation or erosion can produce significant geometrical changes within an specific time interval, reveals the importance of studying the simultaneous effect of the fluid flow, physical or chemical reaction, and geometry on every capillary within the network. Even though every capillary is linked, the complex underlying phenomena are influenced by the local geometry of the rock sample under study. When simulating a specific process, the capillaries under its effects may evolve differently and under vastly different time scales.
When considering the use of catalyst to enhance precipitation, careful consideration should be taken to the rate in order to optimize permeation of the pore space and thus the total volume stored. Transport-reaction simulations like those presented here that allow studying the effect of the injection conditions and reaction parameters on relevant sample sizes may be valuable in selecting the optimum rates.

\section*{Conclusions}\label{sec:conclusions}


We have developed an iterative methodology to simulate transport-reaction processes within capillary network representation of porous media and have investigated the effect of varying parameters (e.g., fluid flow conditions, phasic properties, and rock sample geometry) on the onset of physical and chemical pore-scale processes, and the consequent spatio-temporal evolution in relevant rock sample sizes. The fluid-solid interface evolution due to pore-scale interactions is modeled by estimating changes in the capillary diameter through an iterative computation in time. We have collected expressions from the literature to predict the rates, time scales and impact on the capillary network of several pore-scale processes. 

We studied the effect of precipitation on the geometry evolution and various rock properties like precipitation volume or onset of clogging, assuming a calcium carbonate matrix injected with brine with the geometry extracted from a Berea sandstone rock.
The results indicated that more accurate predictions are possible when accounting for the influence of the local flow and geometry conditions on the material accumulation estimates.

\bibliography{Simulating_carbon_mineralization_at_pore_scale_in_capillary_networks_of_digital_rock}

\section*{Acknowledgements}

The authors acknowledge project support by Bruno Flach and Alexandre Pfeifer (both IBM). Also, we thank Vassilis Vassiliadis, Alessandro Pomponio, and Michael Johnston (all IBM Research) for expert technical assistance.

\section*{Author contributions statement}
 
Authors D.A.L.V., J.T.A., R.G., R.N.B.F. and M.E.F. developed the methodology used by the simulator. 
Author D.A.L.V. implemented the \textit{Fluid-Solid Process Simulator}, \textit{Network Geometry Modifier}, \textit{Model Reaction Library} and data visualization scripts. 
Authors D.A.L.V., J.T.A. and M.F.B.R. performed the simulations, implemented data aggregation and analysis scripts, generated graphs and prepared figures. 
Authors D.A.L.V, R.N.B.F. and M.F.B.R. implemented the integration of the flow simulator with ST4SD. 
Authors D.A.L.V. and J.T.A. wrote the manuscript and supplemental materials. Authors D.A.L.V. and J.T.A., R.G., R.N.B.F., M.E.F., and M.B.S. performed data interpretation and contributed to the manuscript and supplemental materials. All authors reviewed the manuscript.

\section*{Code and data availability}\label{sec:code}

The code used to obtain the CNM representations of the rock samples contained in the repositories is available in a Github repository (\url{https://github.com/IBM/flowdiscovery-digital-rock}). Additional algorithms used for processing and segmenting the raw grayscale images, are available as Python code (\url{https://github.com/IBM/microCT-Dataset}). The code to run flow simulations and Geometry Modification Module on CNM representations of the rock samples is also available as a Github repository (\url{https://github.com/IBM/flowdiscovery-simulator}). Finally, the code to automate the execution of the simulations using the Simulation Toolkit for Scientific Discovery (ST4SD)~\cite{ST4SD} can also be found at Github (\url{https://github.com/st4sd/flow-simulator-experiment}).

\noindent Microtomography datasets containing grayscale and binary rock image data are available at the Digital Rocks Portal for sandstone samples (\url{https://dx.doi.org/10.17612/f4h1-w124}) and on Figshare for sandstone and carbonate samples (\url{https://doi.org/10.25452/figshare.plus.21375565.v6}).

\section*{Additional information}

The authors declare no Competing Financial or Non-Financial Interests.



\section*{Supplementary Information}

\renewcommand{\thesection}{S.\arabic{section}}

\renewcommand{\theequation}{S.\arabic{equation}}
\setcounter{equation}{0}
\renewcommand\thesupptable{S.\arabic{supptable}}
\renewcommand \thesuppfigure{S.\arabic{suppfigure}}

\input{supplemental-models}

\end{document}

%% file: supplemental-models.tex
\subsection*{Automated simulation workflow} \label{sec:ST4SD}  

The large number of scenarios with different simulation parameters lead to a parameter space that is intractable, but by leveraging high-throughput automated simulation workflows. Given the large parameter space that needs to be swept to fully assess the mineralization potential in a given geological storage scenario, we employed the Simulation Toolkit for Scientific Discovery (ST4SD)~\cite{ST4SD} to automate the execution of long simulation campaigns with several chained steps. By doing so, we drastically reduce the human labor involved in preparing the simulation inputs, submitting the jobs, and post-processing the results. The use of such a workflow scheduler also ensures the reproducibility of our results and enables efficiency gains by optimizing the use of computing resources. The underlying code implementation for this automation has been shared for repeatability (see ``Code Availability'' section).

\begin{suppfigure}[h]
\centering
\includegraphics[width=\linewidth]{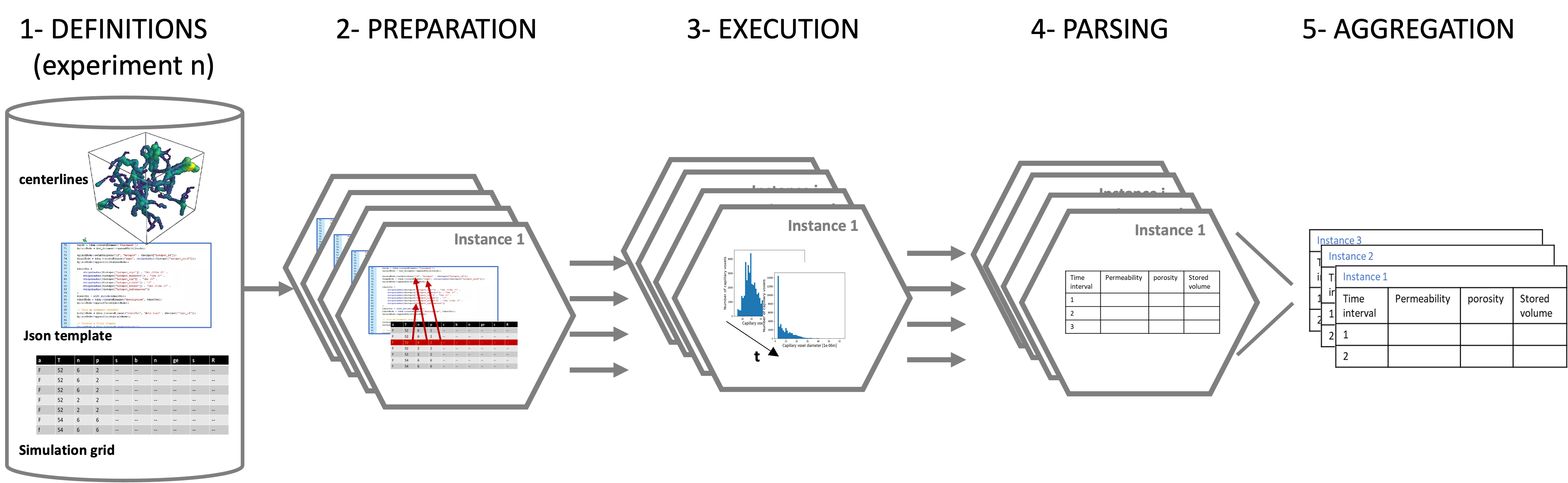}
\caption{Automated computational methodology showing, from left to right, experiment definition, preparation of inputs, simulation execution, output parsing, and aggregation of results.}
\label{fig:ST4SD}
\end{suppfigure}

Figure \ref{fig:ST4SD} shows the sequence of steps executed by each ST4SD experiment comprising 1) an experiment \textit{definition}, needed for the 2) \textit{preparation} of inputs for the 3) \textit{execution} of the simulation, followed by the output 4) \textit{parsing}, leading tot he final 5) \textit{aggregation} of all the results. The user input at the \textit{definition} step comprises three parts: the rock sample CNM in which the simulations are performed; a template of the simulation configuration file where to enter all fluid, rock, and reaction properties (e.g., viscosity, reaction rates), as well as the flow simulation conditions (e.g., pressures, temperatures) necessary for each execution; and a simulation grid where each line defines the values of all properties and conditions, including the types of reactive transport phenomena considered, that are applied to each simulation instance.    The next step is the \textit{preparation}, for each simulation instance, of the configuration files and inputs based on the defined user inputs. Each combination of rock and fluid samples, flow simulation conditions, and reactive transport parameters extracted from the simulation grid leads to a different simulation scenario or instance. Many simulation conditions are not explicitly provided by the user, so default values are used instead. 

The simulations are automatically scheduled and executed by ST4SD workflow, running all scenarios in parallel during the \textit{execution} step. Per instance, this step corresponds to the execution of the coupled reactive-transport simulation, as illustrated in Fig \ref{fig:GMM_workflow}, consisting of the iterative workflow where single-phase flow simulations are intercalated with a geometry modification routine that updates the capillary network based on the reactive phenomena. The \textit{execution} step runs until reaching a convergence criteria and is then followed by the \textit{parsing} of the simulation results to extract the relevant figures-of-merit from each parallel execution. Values of the sample porosity or the accumulated precipitated volume, for instance, are computed after each iteration and then saved into a single document during \textit{parsing}, thus allowing eliminating memory-consuming interim files. A final post-processing of results is performed during the \textit{aggregation} step, where all relevant figures-of-merit from all parallel simulations are collected into a single table format to facilitate user consumption.

\subsection*{Reaction Model Library}\label{sec:SuppLibrary}

Within each iteration of the workflow in Fig. \ref{fig:GMM_workflow}, the \textit{fluid-solid process simulator} module takes as input the fluid phasic properties and flow conditions and pore-scale process parameters defined during the experiment preparation, as well as the flow property fields at that time step in every capillary in the sample, and computes the changes to the capillary diameters according to the reaction models provided by a reaction model library, which comprises methods and correlations for simulating the geometry evolution of porous media due to erosion, dissolution, and precipitation. The inputs and outputs of the model are depicted in Fig. \ref{fig:GMMmethod}a. Any new model or phenomenon formulation that adapt to the current set of input parameters can be readily added to the library. Additional parameters are possible but requires adjustments to the input configuration files.

\subsubsection*{Erosion and Deposition}

Physical pore-scale processes like erosion and deposition are frequently modeled through deterministic approaches\cite{bonelli2006modelling, bonelli2013erosion, hanson2010internal, jager2017channelization, wahl2008erosion, wan2004investigation, wan2009experimental,matias2021flow}. We modeled the effect of erosion on the pore geometry by calculating each capillary diameter variation according to the shear stress thresholds of the erosion and deposition law in Jager et al.\cite{jager2017channelization} and Matias et al. \cite{matias2021flow}. Considering slow erosion and deposition, where the surface change is much slower than the fluid flow velocity, these models predict rates for the evolution of the diameter of spheres. 
The prediction of the threshold for the onset of the geometry modification is based on the assumption of critical shear stresses, which depends on the fluid flow and the phasic characteristics. Before estimating changes to the diameter, this model first computes a threshold for the onset of erosion per capillary. Since the pore-scale processes involved are geometry and flow-dependent, erosion or deposition may not occur in all capillaries within the network as each has different flow characteristics (i.e., flow rate, pressure gradient). As a  consequence, the spatial domain does not follow a uniform evolution in all capillaries (see Fig. \ref{fig:OnsetErosion}a of the distribution of the onset of erosion). The erosion and deposition law \cite{jager2017channelization} reads:
\begin{equation}\label{eq1_thresholds}
    \dot{r}_{er/dep}=\left\{\begin{matrix}
    - \kappa_{er} \left ( \tau_w - \tau_{er} \right ) & \textup{if} \; \tau_w > \tau_{er}\\ 
    C \kappa_{dep} \left ( \tau_{dep} - \tau_w \right ) & \textup{if} \; \tau_w \leq \tau_{dep} \\ 
    0 & \textup{if} \; \tau_{er} \geq \tau_{w} \geq \tau_{dep}
    \end{matrix}\right.
\end{equation}
where $\dot{r}_{er/dep}$ represents the erosion or deposition rate (kg m$^{-2}$), $\kappa_{er}$ the erodibility coefficient, $\kappa_{dep}$ the deposition coefficient, $C=C(t)/C_{0}$ is the relationship between the current and fully saturated concentrations, $\tau_{w}$ represents the wall shear stress, $\tau_{er}$ the critical shear stress above which  erosion occurs, and $\tau_{dep}$ the critical shear stress above which  deposition occurs. The linear relation of erosion to wall shear stress is an empirical law that is well established and reasonable for a variety of materials such as cohesive soils, clay, and materials of similar properties \cite{jager2017channelization, bonelli2006modelling, bonelli2013erosion}. 

The erodibility coefficient depends on pure shear stress, turbulent fluctuations of shear stress, turbulent fluctuations of normal stress \cite{briaud2008case}, and properties of the solid matter (e.g., the cohesive forces) \cite{grabowski2011erodibility}. The density of the removed mass equals the fluid density, and the eroded mass is not tracked once it is dragged by the flow \cite{pereira2016pore}. 

According to the erosion/deposition law in Jager et al.\cite{jager2017channelization}, as the capillary diameter increases, the shear stress increases, which enhances erosion \cite{matias2021flow}. Erosion rate is proportional to the wall shear stress \cite{parker2000purely},  since the wall shear stress decreases with the diameter. Although most coefficients are obtained from laboratory experiments, as in Bonelli et al. \cite{bonelli2006modelling, bonelli2013erosion}, in this study, the correlations and parameter values of Eq. \ref{eq1_thresholds} are adapted for erosion and deposition shear stress thresholds around the wall-shear stress, following the procedures in Jager et al.\cite{jager2017channelization}, Bonelli et al.\cite{bonelli2006modelling, bonelli2013erosion}, and Parker et al. \cite{parker2000purely}. 

For a constant pressure drop distribution across each capillary, the wall \cite{matias2021flow} and erosion \cite{bonelli2006modelling} shear stresses are denoted by the following correlations: 
\begin{equation}\label{eq_wallshear}
    \begin{split}
        \mathbf{\tau_w} (t_{er}) = \frac{\Delta \mathbf{p}}{\mathbf{L}}\mathbf{D}(t_{er}) \\ \\
        \mathbf{\tau}_{er}(t_{er}) = \Delta \mathbf{p} \mathbf{D}(t_{er})/\mathbf{D}_0
    \end{split}
\end{equation}\\
where $\Delta \mathbf{P}$, $\mathbf{L}$, and $\mathbf{D(t)}$ represent the arrays containing the capillary pressure gradients, capillary length, and capillary diameter, respectively,  within the capillary network at time $t_{er}$. The array containing the initial capillary diameters is denoted by $\mathbf{D_{0}}$.

Considering that the simulations reach an erosion time $t_{\dot{r}_{er}}$ necessary to start the process uniformly, the scaling correlation for the capillary diameter evolution due to erosion \cite{jager2017channelization} reads:
\begin{equation}\label{eq_erosiondiameter}
     \mathbf{D}\left ( t_{er} \right ) = \mathbf{D}_0 e^{t_{\dot{r_{er}}}/t_{er}}
\end{equation}
where the erosion timescale $t_{\dot{r}_{er}}$is denoted by $t_{\dot{r_{er}}} = 2 \left ( \rho_{s} / \kappa_{er} \right ) \left ( \mathbf{L} /\Delta p \right )$. 

The scaling correlation for the capillary diameter evolution due to deposition \cite{jager2017channelization} reads:
\begin{equation}
     \mathbf{D} \left ( t_{dep} \right ) = \mathbf{D}_0 \left [ \exp \left ( \frac{\kappa_{dep} \Delta P \, C}{2 \rho_s L} t_{\dot{r}_{dep}} \right ) \right ] + 4 \frac{\tau_{dep} \mathbf{L}}{\Delta P}
\end{equation}
where $t_{\dot{r}_{er}}$ is the necessary time to start the deposition uniformly within a capillary. 

Figure \ref{fig:OnsetErosion}a displays a distribution of capillaries under the effect of erosion (label 1) based on the calculation of reaction onset thresholds\cite{jager2017channelization} with an erodibility coefficient $\kappa_{\textup{er}} = 0.071s\cdot m^{-1}$, 1000 secs of simulation time on a small sandstone rock sample of $225\mu m\times225\mu m\times900\mu m$ in size.

The pore-scale processes involved are geometry and flow dependent, which determine the reaction rate as well as the potential existence of a process within each capillary as determined by Eq. (\ref{eq1_thresholds}), and thus the number of eroded capillaries in the network may vary after each simulation time step.
Moreover, the geometrical variation of each capillary within the network can be different given the varying flow characteristics (i.e., flow rate, pressure gradient) that may lead to either erosion, deposition, or none of those processes. When no process is present, the diameter remains constant.
In the simulation results displayed in Fig. \ref{fig:ErosionEvolution}a, only the portion of the capillaries satisfying the criteria are affected by erosion and thus a non-uniform evolution of the geometrical changes is expected between the time steps t$_0$,t$_{1000}$, and t$_{5500}$, corresponding to instants 0s, 17s and 93.5s, respectively.  Fig. \ref{fig:ErosionEvolution}b shows the distribution of capillary diameters at those same time steps. 
As a consequence, the porosity of the small rock sample displayed in Fig \ref{fig:OnsetErosion}a increases until reaching a plateau as shown in Fig \ref{fig:OnsetErosion}b, where the geometry and flow conditions are not sufficient to meet the erosion and Deposition Law thresholds and, therefore no more variation in the diameter of the capillaries in the network occurs due to erosion.

\begin{suppfigure} [ht]
    \centering
    \includegraphics[width=\linewidth]{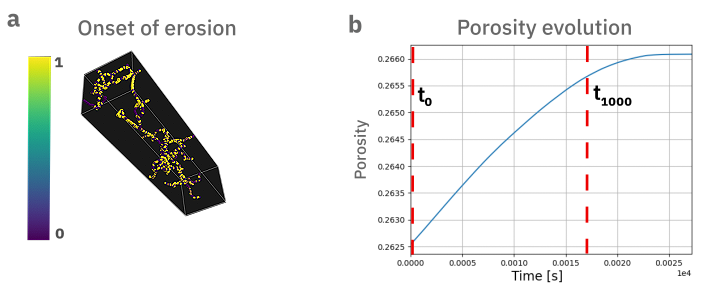}
    \caption{a) Distribution of capillaries under the effect of erosion (label 1) based on the calculation of reaction onset thresholds\cite{jager2017channelization} with an erodibility coefficient $\kappa_{\textup{er}} = 0.071s\cdot m^{-1}$, 1000 secs of simulation time on a small sandstone sample of size $225\mu m\times225\mu m\times900\mu m$.  b) Evolution of porosity as a result of erosion with time.}
    \label{fig:OnsetErosion}
\end{suppfigure}

\begin{suppfigure} [ht]
    \centering
    \includegraphics[width=\linewidth]{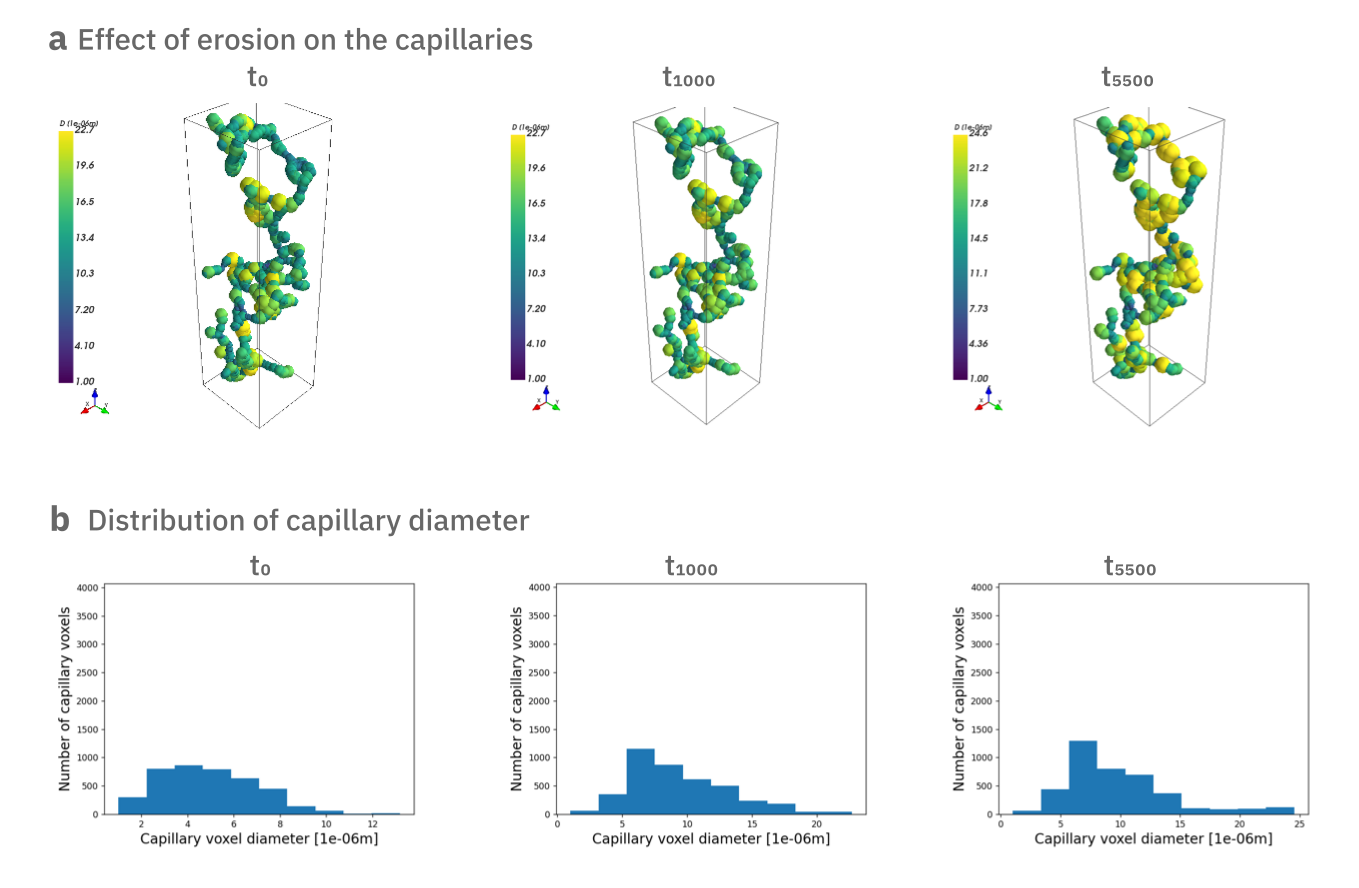}
    \caption{a) Evolution of the capillaries meeting the criteria for erosion at three simulation time steps t$_0$, t$_{1000}$, and t$_{5500}$, corresponding to instants 0s, 17s and 93.5s, respectively. b) Distribution of capillaries diameters as a result of erosion at those same time steps. } 
    \label{fig:ErosionEvolution}
\end{suppfigure}

\subsubsection*{Mineral dissolution}

Several authors have studied the effects of mineral dissolution on the geometry evolution of porous media as a function of the concentration fields. Existing models resolve single \cite{molins2021simulation} and multiphase flow \cite{molins2012investigation, molins2014pore, molins2017mineralogical, maes2021geochemfoam} transport equations combined with reaction equations. Computation of these model on large or complex domains can be costly. 

The dissolution of the rock wall can lead to the release of cations (eg. Mg$^{(2+)}$, Ca$^{(2+)}$, Fe$^{(2+)}$  into solution which will, in turn, react with dissolved CO$_2$ to form stable carbonates. The formation of these new stable chemicals species can lead to the mineral storage of CO$_2$\cite{juncu2020}.
In one example, the dissolution of calcium carbonate contributes to the mass of dissolved components according to the following stoichiometric relationship:
\begin{equation}\label{dissolution}
    CaCO_3 \left ( s \right ) \rightarrow Ca^{2+} + CO_3^{2-}
\end{equation}
The Transition State Theory rate law is used in the calculation of the reaction rates governing the mineral dissolution processes ($r_m$ $\left [ \textup{mol}/\textup{m}^{2}\textup{s} \right ]$). The equations are based on a pore-scale flow and reactive transport model proposed by Li et al. \cite{li2010three} and Molins et al. \cite{molins2021simulation, molins2014pore, molins2015reactive, molins2017mineralogical}. Calcium carbonate dissolution reaction becomes increasingly limited by diffusion\cite{molins2017mineralogical}. The dissolution reaction rate \cite{molins2021simulation} $\dot{r}_{dis}$ in units of mol m$^{-2}$ s$^{-1}$ is denoted by:
\begin{equation}
    \dot{r}_{dis} = k_{H^{+}} \gamma_{H^{+}} c_{H^{+}}
\end{equation}
where $k_{H^{+}}$, $\gamma_{H^{+}}$, and $c_{H^{+}}$ represent the dissolution, thermodynamic activity, and concentration coefficients, respectively. The scaling correlation for the capillary diameter evolution due to mineral dissolution (single-phase flow) reads:
\begin{equation}
    \mathbf{D} \left ( t_{\dot{r}_{dis}} \right ) = \left [ \mathbf{D}_{0}^2 + \frac{\mathbf{D}_0 \dot{r}_{ppt}}{\rho_s} \right ]^{1/2} t_{\dot{r}_{dis}}
\end{equation}

\subsubsection*{Mineral precipitation}

Noiriel et al.\cite{noiriel2012upscaling} proposed an integrated experimental and modeling approach for studying an induced calcium carbonate growth in cylindrical cores packed with glass beads and calcium carbonate crystals injected over 28 days with a supersaturated mixture of CaCl$_{2}$ and NaHCO$_{3}$. A general equation to describe the rate of calcium carbonate precipitation \cite{lasaga2001variation} ($\dot{r}_{ppt}$ $\left [ \textup{mol}/\textup{m}^{2}\textup{s} \right ]$) reads:
\begin{equation} \label{eq:precipitation rate}
    \dot{r}_{ppt} = k_{ppt} \left [ \exp \left ( \frac{m \Delta G}{R^{*}\cdot T} \right )-1 \right ]^{n} = k_{ppt} \left [ \exp\left ( \frac{IAP}{K_{sp}} \right )^{m}-1 \right ]^{n} = k_{ppt} \left ( \Omega^{m}-1 \right )^{n}
\end{equation}

where $k_{ppt}$ represent the precipitation rate constant (mol m$^{-2}$ s $^{-1}$), $\Delta G$ the Gibbs free energy change of the overall reaction (J  mol$^{-1}$), $\Omega$ the saturation index $\Omega = IAP/K_{sp}$, $K_{sp}$ the solubility of calcium carbonate, $IAP$ the ion activity product defined by $a_{Ca^{2+}}a_{CO_3^{2-}}$, with $a_{Ca^{2+}}$ and $a_{CO_3^{2-}}$ the activities of $Ca^{2+}$ and $CO_3^{2-}$, respectively. $R^{*}$ represents the gas constant (J K$^{-1}$ mol$^{-1}$) and $T$ the absolute temperature (K). The values of $n$ and $m$ are semi-empirical constants that depend on the kinetic behavior involved in the chemical reaction. The coefficient $m$ can be interpreted as the Temkin coefficient, relating the reaction stoichiometry in the rate-limiting step to the reaction stoichiometry of the overall reaction \cite{aagaard1982thermodynamic, oelkers1994effect}. Mineral trapping is modeled through an exponential decay model, considering a minimum diameter for the flow simulator to resolve the equation system.

In our work we implemented two approaches for simulating mineral precipitation (i.e., uniform and flow-dependent crystallization) of calcium carbonate rock samples in brine and the consequent clogging in time using the correlations of Lasaga \cite{lasaga1981kinetics} and Noiriel et al.\cite{noiriel2012upscaling}. We analyze the effect of flow conditions and mineral precipitation parameters on the evolution of the accumulated material distribution within capillary networks.

\subsubsection*{Uniform precipitation rate}\label{precipitation_uniform}

In this simpler approach, we track the geometry evolution due to mineral precipitation under the condition of constant and uniform precipitation rates in all capillary within the network using Eq. \ref{eq:precipitation rate}. After each reaction occurs, we consider a uniform decrease in the reactive area within each capillary. The scaling law for the modified capillary diameter, considering uniform precipitation rate in a generic reactive area, phasic properties, and reaction time, reads:
\begin{equation} \label{eq:correlation1}
    \mathbf{D} \left ( t_{\dot{r}_{ppt}} \right ) = \left [ \mathbf{D}_{0}^2 - \frac{\mathbf{D}_0 \dot{r}_{ppt}}{\rho_s} \right ]^{1/2} t_{\dot{r}_{ppt}}
\end{equation}
where $ \mathbf{D}_{0}$ represents the vector or initial capillary diameters, $t_{\dot{r}_{ppt}}$ the precipitation reaction time, and $ \rho_s $ the solid phase density. The precipitation rate $\dot{r}_{ppt}$ and the reaction time $t_{\dot{r}_{ppt}}$ are defined by Eq. \ref{eq:precipitation rate} and the parameters of Tables 2 and 7 in Noiriel, C., et al.\cite{noiriel2012upscaling}. The Model has the capability of setting input parameters from other experimental or numerical methods obtained by other authors.

\subsubsection*{Flow dependent precipitation rate}\label{precipitation_variable}

The material accumulation at each capillary evolves due to the effect of a capillary-dependent flow rate distribution and an input calcium concentration variation obtained from the literature. The modified capillary diameter, obtained from the calculation of the accumulated volume using Eq. (3) in Noiriel et al.\cite{noiriel2012upscaling}, reads:
\begin{equation} \label{eq:correlation2}
     \mathbf{D} \left ( t_{\dot{r}_{ppt}} \right ) = -\frac{\mathbf{Q} \left ( t_{\dot{r}_{ppt}} \right ) \Delta Ca}{\dot{r}_{ppt} \pi \mathbf{L} \left ( t_{\dot{r}_{ppt}} \right )}
\end{equation}
where $\mathbf{Q}$ is the flow rate vector, $\Delta Ca$ is the calcium concentration variation between the inlet and the outlet of each capillary, and $\mathbf{L}$ is the vector containing the capillary length distribution within the network. In a first implementation, we assume a uniform calcium variation concentration $\Delta Ca$ within the capillary network; however, this parameter can be adjusted according to match experimental estimates or values reported in the literature. This method enables comparing the accumulated material under a set of fluid flow, geometry and pore-scale process characteristics.\\

\subsubsection*{Clogging prediction} \label{methods_clogging}

Both physical and chemical mechanisms can cause clogging in porous-media \cite{salles1993deposition, alem2015hydraulic}: (i) Physical clogging, restricting particle movement in soil and other porous media (straining and filtration); and (iii) Chemical clogging, which controls the colloidal stability of the particles and minerals precipitated from water \cite{compere2001transport, mays2007hydrodynamic}. 

Through experimental studies, Alem et al.\cite{alem2015hydraulic} assessed the differences in porous medium physical clogging processes under the condition of constant flow rate and the condition of constant head. Mays and Hunt\cite{mays2007hydrodynamic} presented results that indicate that hydrodynamic aspects of clogging by natural
materials (e.g., Montmorillonite) are consistent with those of simplified model systems. To investigate hydrodynamic effects on clogging, Mays and Hunt \cite{mays2005hydrodynamic} analyzed published data sets from constant-flow filtration experiments that measured particle accumulation and head loss over a range of fluid velocities.